\documentclass[12pt,letter]{article}
\pdfoutput=1

\usepackage{graphicx, epsfig, color,cite}
\usepackage{amsmath}
\usepackage{amssymb}
\usepackage{subfigure}
\usepackage{doi}

\def\eslt{E_T^{\rm miss}}

\def\to{\rightarrow}

\def\bi{\begin{itemize}}
\def\ei{\end{itemize}}

\def\tchi{\tilde\chi}

\def\sps1ap{SPS1a$^\prime$}
\def\c1p{C1$^\prime$}

\def\tst{\tilde t}

\def\tg{\tilde g}

\def\alt{\lesssim}
\def\agt{\gtrsim}
\def\be{\begin{equation}}  
\def\ee{\end{equation}}  
\def\bea{\begin{eqnarray}}  
\def\eea{\end{eqnarray}}  
\def\beas{\begin{eqnarray*}}  
\def\eeas{\end{eqnarray*}}

\begin{document}
\begin{titlepage}
\begin{flushright}
OU-HEP-250830 \\
\end{flushright}

\vspace{0.5cm}
\begin{center}
  {\Large \bf Reach of $e^+e^-$ Higgs factory for\\
    light higgsinos via electroweak precision observables\\
    and comparison with other future facilities
}\\ 
\vspace{1.2cm} \renewcommand{\thefootnote}{\fnsymbol{footnote}}
{\large 
Howard Baer$^{1}$\footnote[1]{Email: baer@ou.edu },
Vernon Barger$^{2}$\footnote[2]{Email: barger@pheno.wisc.edu },
Natsumi Nagata$^{3}$\footnote[3]{Email: natsumi@hep-th.phys.s.u-tokyo.ac.jp}
and Dibyashree Sengupta$^1$\footnote[4]{Email: Dibyashree.Sengupta@roma1.infn.it}
}\\ 
\vspace{1.2cm} \renewcommand{\thefootnote}{\arabic{footnote}}
{\it 
$^1$Department of Physics and Astronomy, \\
University of Oklahoma, Norman, OK 73019, USA \\[3pt]
}
{\it 
$^2$Department of Physics and Astronomy, \\
University of Wisconsin, Madison, WI 53706, USA \\[3pt]
}
{\it 
$^3$Department of Physics, University of Tokyo, Bunkyo-ku, Tokyo 113-0033, 
Japan \\[3pt]
}
{\it
  $^4$ INFN, Sezione di Roma, c/o Dipartimento di Fisica, Sapienza Università di Roma, Piazzale Aldo Moro 2, I-00185 Rome, Italy}\\[3pt]

\end{center}

\vspace{0.5cm}
\begin{abstract}
\noindent 
Light higgsinos with mass $\sim 100-400$ GeV are well-motivated from
naturalness considerations within supersymmetric models.
However, at hadron colliders such as CERN LHC, they are rather difficult to
search for due to the small visible energy release from heavy higgsino decay
to the lightest higgsino, assumed here to be the lightest SUSY particle (LSP).
An alternative way to search for the sparticles of supersymmetry is via their
virtual effects on electroweak precision observables (EWPO) such as the $W$
boson mass or the effective weak mixing angle $\sin^2\theta_{\rm eff}$.
We quantify the ability of an $e^+e^-$ Higgs factory operating at
$\sqrt{s}\sim 90-250$ GeV to indirectly detect higgsinos via EWPO in the
so-called {\it higgsino discovery plane}. The latter allows one to compare
the relative reach of LHC and high-lumi LHC with an $e^+e^-$ Higgs
factory and with a linear $e^+e^-$ collider operating at $\sqrt{s}\sim 0.5$
TeV.
\vspace*{0.8cm}

\end{abstract}

\end{titlepage}

\section{Introduction}
\label{sec:intro}

Supersymmetry (SUSY) is well-motivated in that it provides a 't Hooft
technically natural solution to the gauge hierarchy problem:
it stabilizes the weak scale via the cancellation of otherwise
dangerous quadratic divergences\cite{Witten:1981nf,Kaul:1981wp}.
The quadratic divergence cancellation is maintained under soft SUSY breaking;
the latter is necessary in that no superpartners have so far been experimentally
observed. But there is an expected upper limit on superpartner
masses arising from the so-called {\it practical naturalness principle}\cite{Susskind:1978ms,Veltman:1980mj,Baer:2015rja,Baer:2023cvi}:
{\it all independent contributions to observables should be comparable to or
  less than their measured values}.\footnote{Veltman\cite{Veltman:1980mj} presciently states ``This criterium (of naturalness) is that radiative corrections are supposed to be of the same order (or much smaller) than the actually observed values. And this then is taken to apply also for coupling constants and masses.''}
In the minimal supersymmetric standard model (MSSM)\cite{Baer:2006rs}, the measured value of
the weak scale is related to SUSY Lagrangian parameters via
the scalar potential minimization conditions:
\be
m_Z^2/2=\frac{m_{H_d}^2+\Sigma_d^d-(m_{H_u}^2+\Sigma_u^u)\tan^2\beta}{\tan^2\beta -1}-\mu^2\simeq-m_{H_u}^2-\mu^2-\Sigma_u^u(\tst_{1,2}) .
\label{eq:mzs}
\ee
where $m_{H_u}$ and $m_{H_d}$ are soft breaking SUSY Higgs mass terms,
$\mu$ is the SUSY-conserving Higgs/higgsino mass scale and the
$\Sigma_{u,d}^{u,d}$ contain an assortment of loop corrections
that go as $\Sigma_{u,d}^{u,d}\sim \lambda^2 m_{\rm sparticle}^2/16\pi^2$
where $\lambda$ is some associated gauge or Yukawa coupling\cite{Baer:2012cf}.
The $\tan\beta\equiv v_u/v_d$ is the usual ratio of Higgs field
vacuum expectation values (vevs).
A numerical measure of practical naturalness comes from the
{\it electroweak} finetuning measure defined as\cite{Baer:2012up}
\be
\Delta_{\rm EW}\equiv \mathrm{max}_i| {\it i}\text{-th term on RHS of  Eq.}~\eqref{eq:mzs} |/(m_Z^2/2) .
\label{eq:dew}
\ee

A value $\Delta_{\rm EW}<30$ requires all terms on the right-hand-side (RHS) of
Eq.~\eqref{eq:mzs} to lie within a factor 4 of $m_Z$. 
Eq.~\eqref{eq:dew} is the most conservative measure of naturalness and
is non-optional (as long as $\mu$ is independent of the soft SUSY breaking terms).
Several aspects of natural SUSY models can be read off from Eq.~\eqref{eq:dew}.
\bi
\item Most importantly, the magnitude of $\mu$, which feeds mass to both Higgs and higgsino states, must be in the vicinity of the weak scale: $\mu\sim 100-400$ GeV.
This immediately implies the existence of light higgsinos with mass not too far from the weak scale
given by $m_{\rm weak}\sim m_{W,Z,h}\sim 100$ GeV.
\item The soft term $m_{H_u}^2$ should be driven radiatively to small negative
values $\sim -(100-400)^2$ GeV$^2$ (radiatively-driven naturalness)\cite{Baer:2012cf}.
\item The top-squark contributions to the weak scale are suppressed by loop 
factors and so can lie in the few TeV range, beyond present LHC top-squark 
search limits. The contributions $\Sigma_u^u(\tst_{1,2})$ actually decrease for
large $A_t$ values, which also acts to uplift $m_h\to  \sim 125$ GeV\cite{Baer:2012up}.
\item First and second generation squarks and sleptons can live in the $10-50$ TeV
range since their contributions to $m_{\rm weak}$ are Yukawa-suppressed. Such large
values allow for a mixed quasi-degeneracy/decoupling solution to the SUSY flavor
and CP problems in gravity-mediation\cite{Dine:1990jd,Cohen:1996vb,Baer:2019zfl}.
\ei

From the above considerations, it seems likely that the higgsinos should be
the lightest, or among the lightest, superpartners. Thus, the search for natural
weak scale SUSY may be best served by the search for light higgsino pair production
at hadron colliders like the CERN LHC\cite{Baer:2011ec}. 
The lightest higgsino, $\tchi_1^0$, 
is likely inert and would escape detection at collider experiments, leading to the
appearance of missing $E_T$ ($\eslt$).\footnote{The lightest higgsino could be completely inert under $R$-parity conservation or inert on the timescale of collider detectors under $R$-parity violation induced by higher-order operators\cite{Baer:2025oid}.}
Owing to the expected small mass gap between the heavier ($\tchi_2^0$ and
$\tchi_1^\pm$) and lightest higgsinos, the heavier higgsinos decay via three-body 
modes $\tchi_2^0\to f\bar{f}\tchi_1^0$ and $\tchi_1^\pm\to f\bar{f}^\prime\tchi_1^0$
leading to very soft visible energy. 
The small visible energy release from higgsino pair production makes higgsino detection at collider experiments very challenging.
Nonetheless, higgsino pair production can be searched for in the case where the higgsino
pairs recoil against a hard initial QCD radiation, which also provides for a detector trigger\cite{Han:2014kaa,Baer:2014kya}. 
Search results are typically plotted in the $m_{\tchi_2^0}$ vs. $\Delta m^0\equiv m_{\tchi_2^0}-m_{\tchi_1^0}$ {\it higgsino discovery plane}\cite{Baer:2020sgm}.
Both ATLAS\cite{ATLAS:2019lng} and CMS\cite{CMS:2021edw} in fact report some $\sim 2\sigma$
excesses in these channels from Run 2 data.  Results from LHC Run 3 
and high-luminosity (HL-)LHC (with possibly improved cuts\cite{Baer:2021srt}) are eagerly anticipated. 
But even with the full integrated luminosity of HL-LHC, the entire natural portion
of the higgsino discovery plane does not appear to be accessible to 
future LHC searches\cite{Baer:2021srt}.

Going beyond HL-LHC\cite{Baer:2025zqt}, a linear $e^+e^-$ collider should have no problem 
discovering light higgsinos provided that $\sqrt{s}\agt 2m(\text{higgsino})$\cite{Baer:2014yta,Baer:2019gvu}.
At the present time, it is unclear whether such a machine shall be built\cite{LinearColliderVision:2025hlt}.

An alternative is to build an $e^+e^-$ Higgs factory such as 
FCC-ee\cite{Andre:2025bpv, FCC:2025lpp} or CEPC\cite{CEPCStudyGroup:2018ghi} with $\sqrt{s}\sim 90-250$ GeV. 
The main goal of such a machine would be to make precision measurements of Higgs boson properties such as mass and branching fractions.\footnote{For prospects of Higgs precision measurements in probing SUSY with radiatively-driven naturalness, see Ref.~\cite{Bae:2015nva}.}
Such a machine would also be adept at precision electroweak observables (EWPO) 
such as improved measurements of SM gauge couplings, $m_{W,Z}$ and the 
effective Weinberg angle $\sin^2\theta_{\rm eff}$. 
Precision measurements of the latter quantities would be sensitive to virtual effects of new particles~\cite{Knapen:2024bxw, Allwicher:2024sso, Gargalionis:2024jaw, Maura:2024zxz} including the light higgsinos of natural SUSY as explored recently in
Refs.~\cite{Nagata:2025ycf, Greljo:2025ggc}.\footnote{For previous studies on SUSY contributions to EWPOs, see Refs.~\cite{Grifols:1984xs, Barbieri:1989dc, Drees:1990dx, Drees:1991zk, Barbieri:1991qp, Maksymyk:1993zm, Chankowski:1993eu, Pierce:1996zz, Cho:1999km, Heinemeyer:2004gx, Martin:2004id, Marandella:2005wc, Heinemeyer:2006px, Heinemeyer:2007bw}. }

Our goal in the present paper is to translate the reach of an $e^+e^-$ Higgs factory 
for the higgsinos of natural SUSY via EWPO into the same higgsino discovery plane as used
by LHC and its upgrades, and also by linear $e^+e^-$ colliders. 
Such a comparative study allows one to assess which of the present and future facilities
have the best prospects for discovery of light higgsinos. It is also then possible
to compare with theoretical predictions as to where such a signal is most likely to lie
in the higgsino discovery plane. 

On the theory side, we assume SUSY arises from a string flux compactification\cite{Douglas:2006es} within the confines of the string landscape. 
The string landscape scenario\cite{Susskind:2003kw} is supported by the experimental discovery of dark energy\cite{Peebles:2002gy}
(in the form of the cosmological constant?) with a value in accord with Weinberg's
anthropic prediction\cite{Weinberg:1987dv} of $\Lambda_{CC}$. 
Weinberg's scenario received theoretical support
by the discovery of the enormous number of string vacua\cite{Ashok:2003gk} which emerge from string flux compactifications\cite{Bousso:2000xa}. 
In such a scenario, our universe is but one ``pocket universe'' within the greater multiverse\cite{Linde:2015edk}. 
If we restrict ourselves to string vacua which give rise to the MSSM as the low energy effective field theory (EFT)\cite{Arkani-Hamed:2005zuc}, then one expects the individual soft terms\cite{Baer:2020vad} to scan in the multiverse according to a power-law statistical selection factor\cite{Douglas:2004qg,Susskind:2004uv,Arkani-Hamed:2005zuc} which favors larger soft term values over smaller ones. 
However, the derived value of the weak scale should lie within the ABDS window of values\cite{Agrawal:1997gf}
\be
m_Z/2\alt m_Z^{PU}\alt 4m_Z \ \ \ (\text{ABDS window})
\label{eq:abds}
\ee
(here, $PU$ stands for pocket-universe value of $m_Z$ within the greater multiverse) 
so that complex nuclei arise which are necessary for atoms to form 
(atomic principle).
One may then make statistical predictions for sparticle and Higgs boson masses based on what Douglas calls {\it stringy naturalness}\cite{Douglas:2012bu}. 
Such statistical predictions have met with success in that they predict $m_h\sim 125$ GeV with sparticle masses typically beyond present LHC mass limits\cite{Baer:2017uvn}.

Thus, in Sec. \ref{sec:ewpo}, we review briefly how sparticle contributions to
$m_W$ and $\sin^2\theta_{\rm eff}$ arise. In Sec. \ref{sec:reach}, we plot the reach of an $e^+e^-$ Higgs factory for light higgsinos via virtual effects, and compare to the
reach of LHC and its upgrades, and also to direct higgsino pair searches at 
ILC with $\sqrt{s}\sim 0.5$ TeV. A summary and conclusions are given in Sec. \ref{sec:conclude}.

\section{Review of reach for light higgsinos from EWPO}
\label{sec:ewpo}

The dominant contributions of light higgsinos to electroweak precision observables arise from vacuum polarization diagrams of electroweak gauge bosons, which are encapsulated in the oblique parameters~\cite{Peskin:1991sw, Maksymyk:1993zm, Barbieri:2004qk}
\begin{align}
  \hat{S} &\equiv - \frac{\cos \theta_W}{\sin \theta_W} \Pi^{\prime}_{W^3B} (0)  ~, \\ 
  \hat{T} &\equiv \frac{1}{m_W^2} \left[ \Pi_{W^+W^-} (0) - \Pi_{W^3W^3} (0) \right] ~, \\ 
  W &\equiv -\frac{m_W^2}{2} \Pi^{\prime \prime}_{W^3W^3} (0)  ~, \\ 
  Y &\equiv -\frac{m_W^2}{2} \Pi^{\prime \prime}_{BB} (0)  ~, 
\end{align} 
where \(\theta_W\) is the weak mixing angle, and the functions \(\Pi_{VV'} (q^2) \) are defined in terms of the vacuum polarization functions of the electroweak gauge bosons
\begin{equation}
    \Pi^{\mu\nu}_{VV'} (q) = \Pi_{VV'} (q^2) \eta^{\mu\nu} - \Delta_{VV'} (q^2) q^\mu q^\nu ~,
\end{equation}
for $V, V' = W^\pm, W^3, B$.  For the calculation of the light higgsino contribution to the vacuum polarization functions, see Refs.~\cite{Nagata:2025ycf, Martin:2004id, Pierce:1996zz}. In terms of these oblique parameters, the corrections in \( m_W \) and \( \sin^2 \theta_{\mathrm{eff}} \)  are expressed as 
\begin{align}
    \frac{\Delta m_W^2}{m_W^2} &\simeq - \frac{2 \sin^2 \theta_W}{\cos 2 \theta_W} \hat{S} + \frac{\cos^2 \theta_W}{\cos 2\theta_W} \hat{T} 
    + \frac{\sin^2 \theta_W}{\cos 2 \theta_W} W + \frac{\sin^2 \theta_W}{\cos 2 \theta_W} Y ~, \label{eq:delmw}\\ 
    \frac{\Delta \sin^2 \theta_{\mathrm{eff}}}{\sin^2 \theta_{\mathrm{eff}}} &\simeq \frac{1}{\cos 2 \theta_W} \hat{S} - \frac{\cos^2 \theta_W}{\cos 2 \theta_W} \hat{T} - \frac{\sin^2 \theta_W}{\cos 2 \theta_W} W - \frac{\cos^2 \theta_W}{\cos 2\theta_W} Y ~.\label{eq:delsw}
\end{align}
The chargino and neutralino contributions to the oblique parameters are well approximated by~\cite{Marandella:2005wc}
\begin{align}
    \hat{S} &= \frac{g^2 m_W^2}{48\pi^2 M_2^2} \biggl[
    \frac{r (r-5-2r^2)}{(r-1)^4} + \frac{1-2r + 9r^2 -4r^3 +2r^4}{(r-1)^5} \ln r     
    \biggr] 
    \nonumber \\
    &+ \frac{g^2 m_W^2}{96\pi^2 M_2 \mu} \biggl[
        \frac{2 -19r + 20 r^2 -15 r^3}{(r-1)^4} 
        + \frac{2 + 3r -3r^2 + 4r^3}{(r-1)^5} 2r\ln r 
    \biggr]\sin 2\beta ~, \label{eq:shat_app} \\ 
    \hat{T} &= \frac{g^2 m_W^2}{192\pi^2 M_2^2} \biggl[
    \frac{7r - 29 + 16r^2}{(r-1)^3} + \frac{1 + 6r - 6r^2}{(r-1)^4} 6 \ln r    
    \biggr] \cos^2 2 \beta ~, \label{eq:that_app} \\ 
    Y &= \frac{g^{\prime 2} m_W^2}{120\pi^2 \mu^2} ~, \\
    W &= \frac{g^2}{120\pi^2} \biggl[\frac{m_W^2}{\mu^2} + \frac{2m_W^2}{M_2^2}\biggr] ~, \label{eq:w_app} 
\end{align}
where $r \equiv \mu^2/M_2^2$, $M_2$ is the wino mass, and $g$ and $g^\prime$ are the $\mathrm{SU}(2)_L$ and $\mathrm{U}(1)_Y$ gauge coupling constants, respectively.

\begin{table}[t]
  \centering
  \small
  \begin{tabular}{lcccccc}
  \hline \hline
  Observables & Current & HL-LHC & ILC 250 & Giga-$Z$ & CEPC & FCC-ee \\
  \hline 
    $\Delta m_W$~[MeV]  & 13.3~\cite{ParticleDataGroup:2024cfk} & 9.3~\cite{ATLAS:2018qzr} & 2.5~\cite{ILCInternationalDevelopmentTeam:2022izu} &&0.5~\cite{CEPCPhysicsStudyGroup:2022uwl}& 0.18\,(0.16)~\cite{FCC:2025lpp} \\ 
    $\Delta \sin^2 \theta_{\mathrm{eff}}$ [$10^{-6}$] & 130~\cite{ParticleDataGroup:2024cfk} & 150~\cite{ATLAS:2018qvs, CMS:2017vxj} & 23\,(15)~\cite{Mizuno:2022xuk, ILCInternationalDevelopmentTeam:2022izu} & 4.0~\cite{ILCInternationalDevelopmentTeam:2022izu}&1.9~\cite{CEPCPhysicsStudyGroup:2022uwl}& 1.2\,(1.2)~\cite{FCC:2025lpp} \\ 
   \hline \hline
  \end{tabular}
  \caption{The current and future prospects for the precision of EWPOs. The values outside (inside) parentheses indicate statistical (systematic) errors.
  }
  \label{tab:observables}
\end{table}

The current experimental values and uncertainties of $m_W$ and $\sin^2 \theta_{\mathrm{eff}}$ are given by~\cite{ParticleDataGroup:2024cfk}
\begin{equation}
  m_W = 80.3692(133)~\mathrm{GeV}, \quad \sin^2 \theta_{\mathrm{eff}} = 0.23149(13).
\end{equation}
Table~\ref{tab:observables} summarizes these uncertainties along with projected precisions at future experiments.

For the HL-LHC, we adopt the projected precision on $m_W$ for $|\eta_\ell| < 4$ and 200~~pb$^{-1} $ of data~\cite{ATLAS:2018qzr}, and use ATLAS projections for $\Delta \sin^2 \theta_{\mathrm{eff}}$ based on PDF4LHC15$_\text{HL-LHC}$~\cite{ATLAS:2018qvs, AbdulKhalek:2018rok}, with similar sensitivity expected from CMS~\cite{CMS:2017vxj}. At the ILC with $\sqrt{s} = 250~\mathrm{GeV}$, $m_W$ can be measured with a precision of 2--2.5~MeV by combining multiple methods~\cite{ILCInternationalDevelopmentTeam:2022izu, Wilson:2016hne}, while $\sin^2 \theta_{\mathrm{eff}}$ can be determined via the radiative return process. Giga-Z, the ILC run at the $Z$ pole, is expected to improve the precision of $\sin^2 \theta_{\mathrm{eff}}$ by an order of magnitude, reaching $\Delta \sin^2 \theta_{\mathrm{eff}} = 4.0 \times 10^{-6}$ with 100 fb$^{-1}$ per helicity configuration~\cite{ILCInternationalDevelopmentTeam:2022izu}. Circular colliders offer even higher precision. CEPC is expected to reach $\Delta m_W = 0.5$~MeV and $\Delta \sin^2 \theta_{\mathrm{eff}} = 1.9 \times 10^{-6}$~\cite{CEPCPhysicsStudyGroup:2022uwl}, while FCC-ee projections are taken from the FCC Feasibility Study Report~\cite{FCC:2025lpp}.

\section{Reach of an $e^+e^-$ Higgs factory for
  light higgsinos via EWPO}
\label{sec:reach}

In this Section, we investigate the reach of $e^+e^-$ Higgs factories for 
the light higgsinos of natural SUSY within the context of two well-motivated
SUSY models: the two-extra-parameter non-universal Higgs model (NUHM2)\cite{Ellis:2002wv,Baer:2005bu}
and the generalized\cite{Baer:2016hfa} mirage mediation\cite{Choi:2005ge} model (GMM) wherein there are comparable
gravity-mediated and anomaly-mediated contributions to soft SUSY breaking terms.

\subsection{NUHM2 model}

The four-extra-parameter non-universal Higgs model (NUHM4) is defined by the 
parameter space
\be
m_0(1,2,3),\ m_{1/2},\ A_0,\ \tan\beta,\ m_{H_u},\ m_{H_d}
\ee
The NUHM4 model is expected to be a valid representation of how
supergravity from string theory would manifest itself. 
This is because supergravity generically leads to non-universal scalar masses\cite{Soni:1983rm,Hall:1985dx,Kaplunovsky:1993rd,Brignole:1993dj} (giving
rise to the SUSY flavor problem in SUGRA), although intra-generational 
scalar mass universality is well-motivated in that all matter scalars of each 
generation are unified in the 16-dimensional spinor of $SO(10)$. Thus, in SUGRA
one expects $m_{H_u}^2\ne m_{H_d}^2\ne m_0(1)\ne m_0(2)\ne m_0(3)$. 
Actually, the landscape tends to pull both $m_0(1)$ and $m_0(2)$ to 
generation-independent upper bounds in the 30--50~TeV range thus giving rise to
a mixed decoupling/quasi-degeneracy solution to the SUSY flavor
(and CP\cite{Dugan:1984qf}) problems.
Here, for simplicity we just assume all the $m_0(i)\ (i=1-3)$ are unified since the
heavy first/second generation scalar aren't expected to affect the higgsino discovery plane. This places us instead into the two-extra-parameter NUHM2 model.
In NUHM$i$ models ($i=2-4$), it is convenient to sometimes trade the GUT scale parameters
$m_{H_u}$ and $m_{H_d}$ for the corresponding weak scale parameters $\mu$ and $m_A$
via the electroweak symmetry breaking minimization conditions.

For theory reference, we assume a linear draw to large soft terms ($m_{\rm soft}^1$) 
from the string landscape in the NUHM2 model, tempered by the anthropic 
condition that the derived value of weak scale lies within the ABDS window\cite{Agrawal:1997gf} of values Eq.~\eqref{eq:abds}.

\begin{figure}[tbp]
\begin{center}
 \includegraphics[clip, width = 0.9 \textwidth]{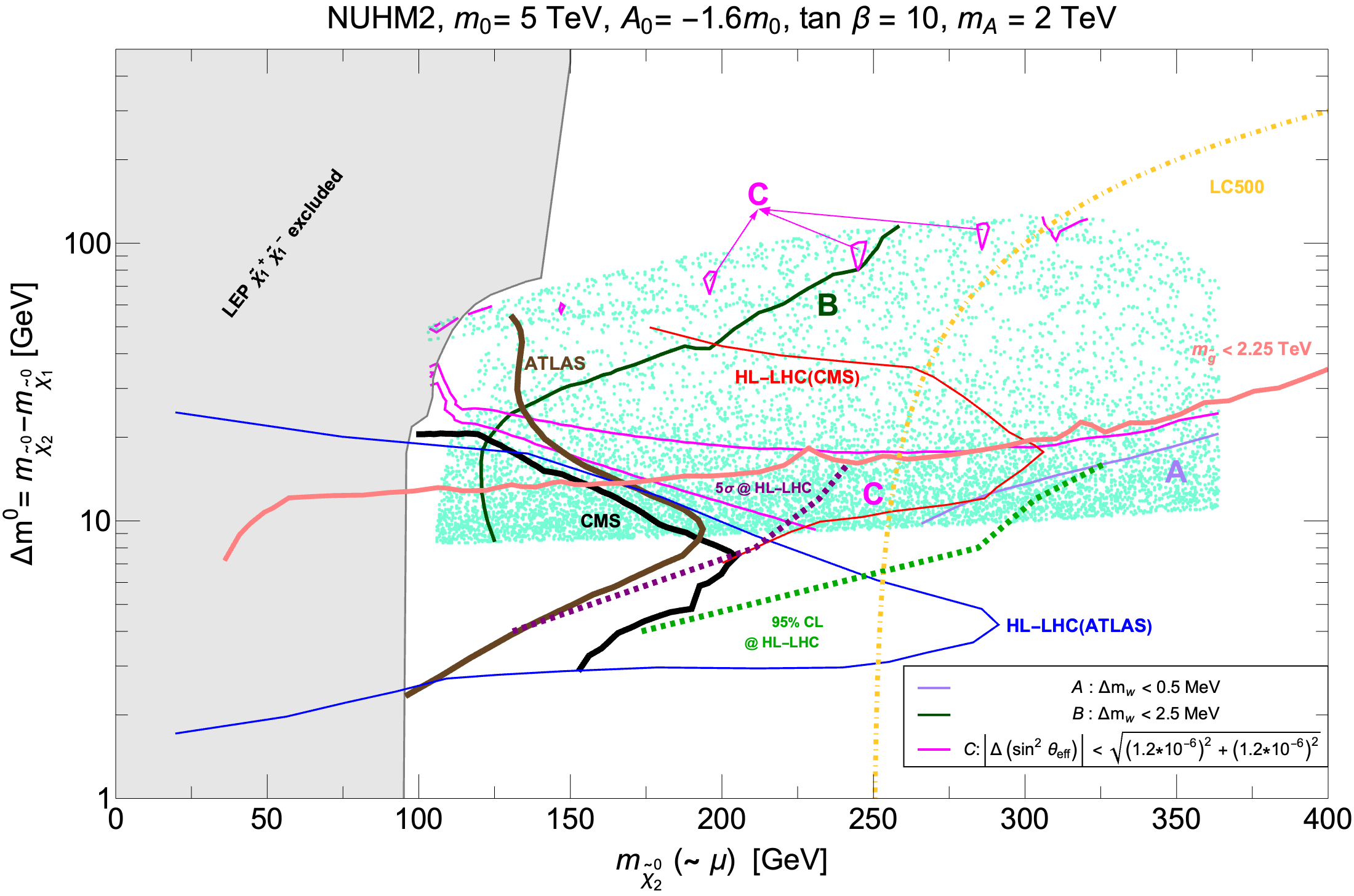}
 \caption{Contours of EWPOs $\Delta m_W$ and $\Delta\sin^2\theta_{\rm eff}$
   in the higgsino mass $m_{\tchi_2^0}\sim \mu$ vs. higgsino mass
   splitting $m_{\tchi_2^0}-m_{\tchi_1^0}$ discovery plane.
   The density of dots reflects  the more stringy natural portion of
   parameter space from landscape selection in the natural NUHM2 
   model. 
   We also show the present reach of the ATLAS experiment and expected
   reach of ATLAS and CMS at HL-LHC and their expected reach via
   implementation of new angular cuts.
   And we show the expected reach of an $e^+e^-$ collider with
   $\sqrt{s}=0.5$ TeV. The region above the $m_{\tg}=2.25$ TeV line
   is excluded by ATLAS/CMS Run 2 gluino pair searches within the framework of simplified models.
   }
\label{fig:NUHM2}
\end{center}
\end{figure}

The NUHM2 dots are for an $n=1$ power-law draw to large $m_{1/2}:0-2$ TeV
with $\mu$ scanned linearly $\mu\sim 100-360$ GeV and with $m_0=5$ TeV, $A_0=-1.6m_0$, 
$\tan\beta =10$ and $m_A=2$ TeV. 
In Fig. \ref{fig:NUHM2}, we plot dots of the two-extra-parameter
non-universal Higgs model (NUHM2) in the higgsino discovery plane,
with a greater density of dots corresponding to greater stringy naturalness.
Along the $x$-axis, the $\tchi_2^0$ mass is $\simeq \mu$, while along the $y$-axis
the larger mass gaps correspond to smaller gaugino masses $m_{1/2}$ (where 
larger gaugino-higgsino mixing leads to larger mass gaps between the two
lightest neutralinos).
The gray region of Fig. \ref{fig:NUHM2} is excluded by LEP2 bounds that
$m_{\tchi_1^\pm}\agt 103.5$ GeV. 
The region {\it above} the pink contour is excluded
by LHC gluino pair searches and has $m_{\tg}<2.25$ TeV. 
We also show the solid ATLAS\cite{ATLAS:2019lng} (brown) and CMS\cite{CMS:2021edw} (black) exclusion contours from
searches for higgsino pair production with a signature of two or three soft leptons 
recoiling against a hard QCD jet plus $\eslt$.\footnote{As noted previously, both ATLAS and CMS seem to have $\sim 2\sigma$ excesses in the soft opposite-sign dilepton $+\text{jet}+\eslt$ channel.} 
Along with these exclusions based on LHC Run 2 data, we also show the projected reach of HL-LHC\cite{Canepa:2020ntc} as computed by 
CMS (red) and ATLAS (blue) and by Ref. \cite{Baer:2021srt} (dashed) using improved angular cuts (green-dashed (95\% CL exclusion) and purple-dashed ($5\sigma$) assuming 3 ab$^{-1}$
of integrated luminosity. 
The HL-LHC projected reaches cover only a portion of the theory-preferred discovery plane.

Relevant to this study, we show contours of 
$\Delta\sin^2\theta_{\rm eff}=\pm 1.7\times 10^{-6}$, a possible goal for EWPO at an $e^+e^-$ Higgs factory such as  FCC-ee at $\sqrt{s}= m_Z$ (solid magenta contours labeled as $C$).
The region to the lower-left of magenta has $\Delta\sin^2\theta_{\rm eff} <-1.7\times 10^{-6}$
while the region to upper-right has positive values. 
Only the gap labeled as C 
between the contours and the holes around $\Delta m^0\sim 100$ GeV (in the LHC excluded region) have tiny $\Delta\sin^2\theta_{\rm eff}$ values.

To the left of the solid green curve 
(labeled $B$) we expect deviations $\Delta m_W$ larger than 2.5 MeV which can be probed by ILC250 (Table \ref{tab:observables}). 
This region is already excluded  by LHC searches for gluinos and by higgsino pair production via soft opposite-sign dleptons, jets$+\eslt$. 
The region to the left of the solid purple curve (labeled $A$), we
expect deviations in $\Delta m_W\agt 0.5$ MeV which could be probed by CEPC. 
Probing deviations all the way to $\sim 0.2$ MeV would allow FCC-ee to explore the remaining lower-right theory-preferred  parameter space beyond the capabilities of HL-LHC.
We also show for comparison the reach of ILC500 for direct charged higgsino pair production $e^+e^-\to \tchi_1^+\tchi_1^-$ as the yellow dot-dashed curve. 
An ILC with $\sqrt{s}\agt 2m(\text{higgsino})\agt 700$ GeV should be able to discover natural SUSY\cite{Baer:2014yta,Baer:2019gvu}.

\begin{figure}[tbp]
\begin{center}
 \includegraphics[clip, width = 0.8 \textwidth]{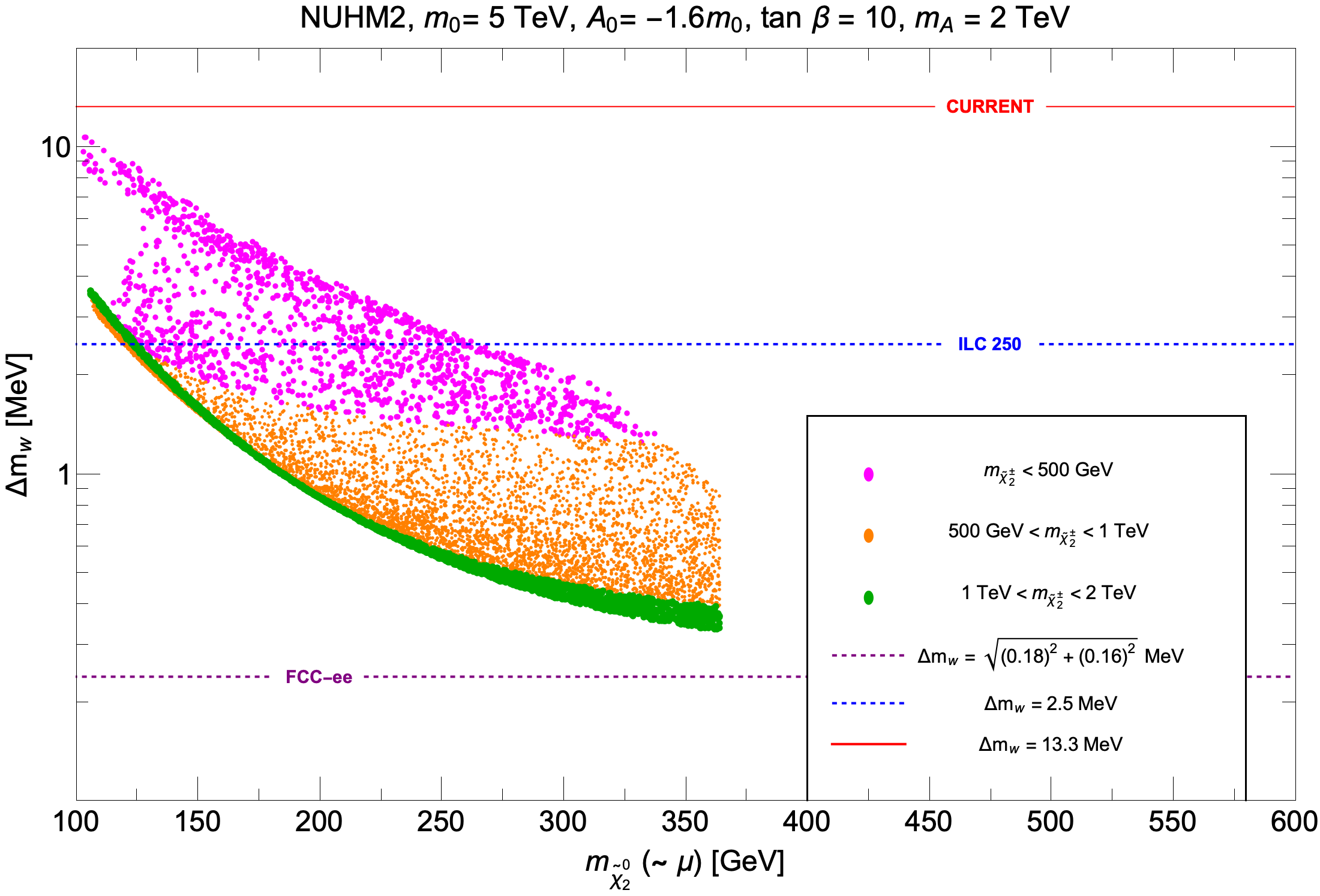}\\
\includegraphics[clip, width = 0.8 \textwidth]{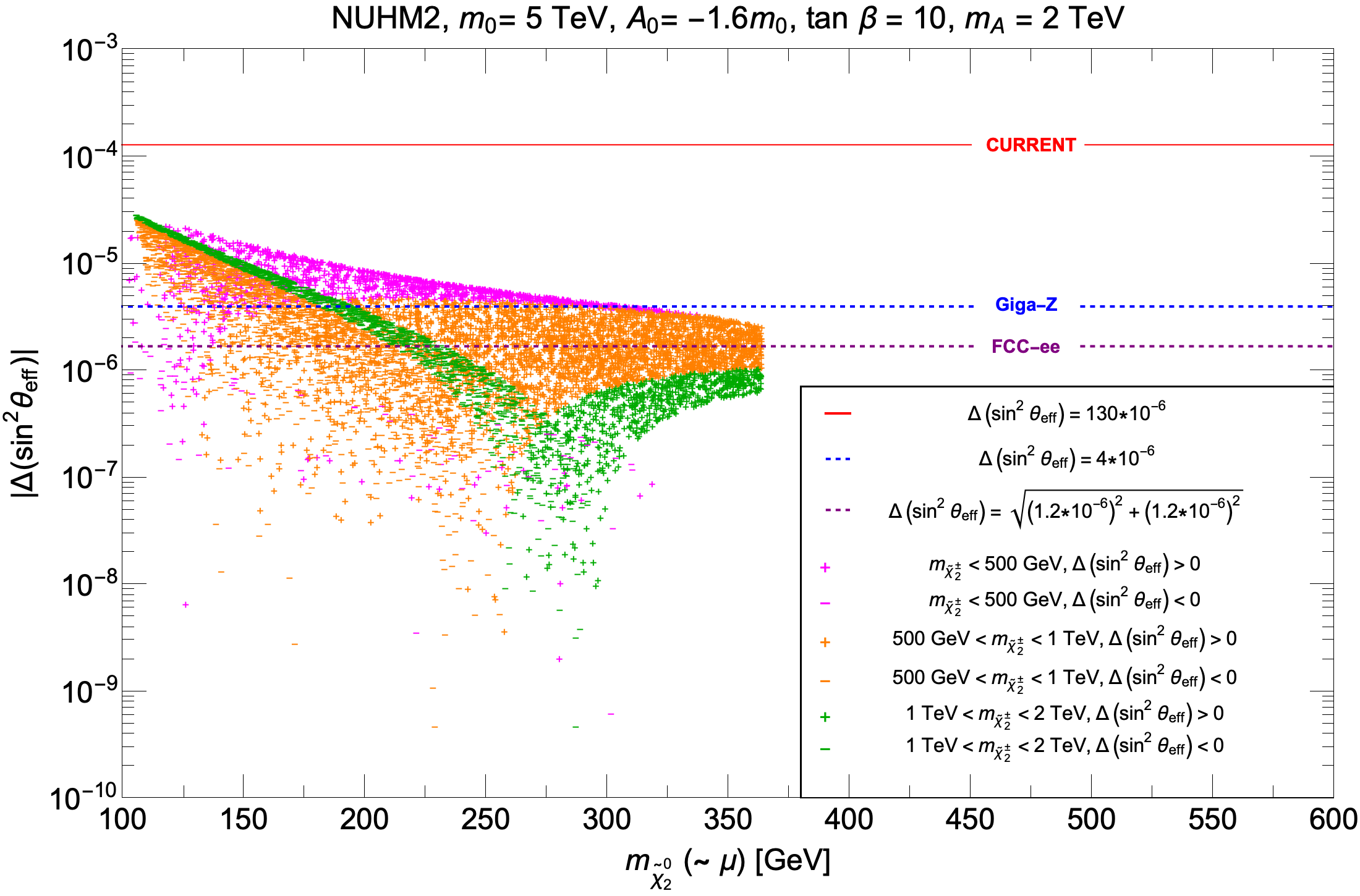}
 \caption{Plot of {\it a}) $\Delta m_W$ and {\it b}) $\Delta\sin^2\theta_{\rm eff}$ 
 vs. $m_{\tchi_2^0}$ from a scan over 
 parameter space of the NUHM2 model with $\mu :100-360$ GeV (uniform distribution) and $m_{1/2}:0-2$ TeV (linear draw to large soft term) and other parameters as shown.
 }
\label{fig:nuhm_scan}
\end{center}
\end{figure}

In Fig. \ref{fig:nuhm_scan}, we show the same landscape theory dots in the {\it a}) $\Delta m_W$ vs. $m_{\tchi_2^0}$ and {\it b}) $\Delta\sin^2\theta_{\rm eff}$ 
vs. $m_{\tchi_2^0}$ planes. 
Again, greater density of dots corresponds to greater stringy naturalness. 
In frame {\it a}), the magenta dots have 
charged wino mass $m_{\tchi_2^\pm}<500$ GeV, orange dots have
$500$ GeV$<m_{\tchi_2^\pm}<$ 1000 GeV and green dots have
1 TeV$<m_{\tchi_2^\pm}<$2 TeV. We also show the present reach lines of
$\Delta m_W=13.3$ MeV and projected future reach of ILC250 of $\Delta m_W=2.5$ MeV and FCC-ee which presumably can probe to $\Delta m_W\sim 0.24$ MeV.
From the plot, the theory points with smaller higgsino or wino masses give the largest
deviations to $m_W$ while the large masses give the expected 
decoupling with smaller contributions to $\Delta m_W$. 
The density of theory points is greatest 
in the decoupling region at the lowest values of $\Delta m_W$.

In Fig. \ref{fig:nuhm_scan}{\it b}), we show the deviations in $\sin^2\theta_{\rm eff}$ vs. $m_{\tchi_2^0}\sim \mu$. 
Here, the values range over positive and negative values 
with positive labeled as $+$s and negative labeled as $-$s.
The largest deviations correspond again to lighter higgsino and wino masses while small deviations correspond to the decoupling case with large sparticle masses, which is preferred by the landscape.

\subsection{Natural generalized mirage mediated SUSY breaking}

\begin{figure}[tbp]
\begin{center}
 \includegraphics[clip, width = 0.9 \textwidth]{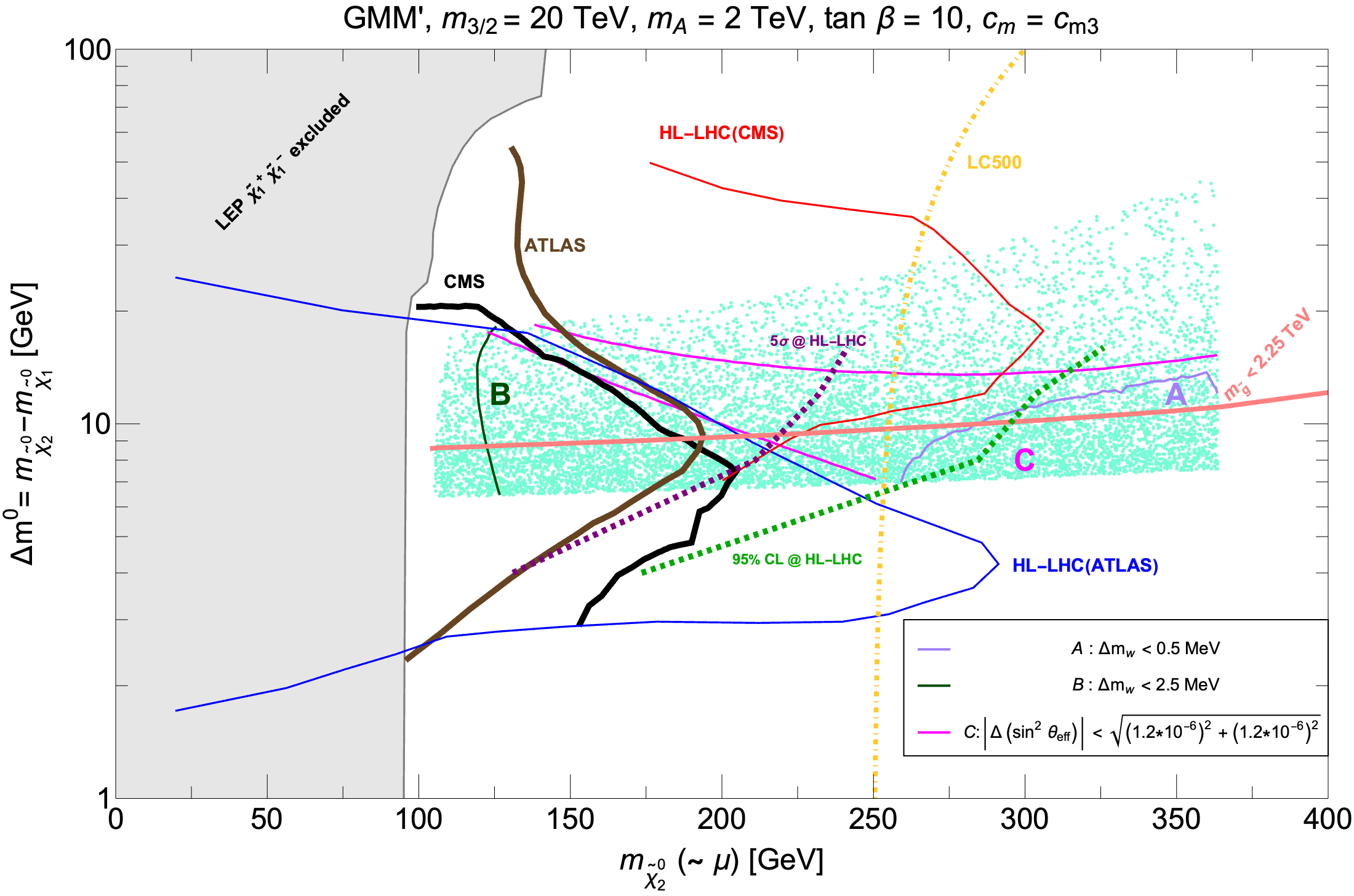}
 \caption{Contours of EWPOs $\Delta m_W$ and $\Delta\sin^2\theta_{\rm eff}$
   in the higgsino mass $m_{\tchi_2^0}\sim \mu$ vs. higgsino mass
   splitting $m_{\tchi_2^0}-m_{\tchi_1^0}$ discovery plane.
   The density of dots reflects  the more stringy natural portion of
   parameter space from landscape selection in natural
   generalized mirage mediation (GMM$^\prime$ model).
   We also show the present reach of the ATLAS experiment and expected
   reach of ATLAS and CMS at HL-LHC and their expected reach via
   implementation of new angular cuts.
   And we show the expected reach of an $e^+e^-$ collider with
   $\sqrt{s}=0.5$ TeV. The region above the $m_{\tg}=2.25$ TeV line
   is excluded by ATLAS/CMS Run 2 gluino pair searches within the framework of
   simplified models.
   }
\label{fig:GMM}
\end{center}
\end{figure}

In Fig. \ref{fig:GMM}, we plot dots corresponding to a landscape scan over
the natural generalized mirage mediation (nGMM) SUSY breaking model\cite{Baer:2016hfa}, 
where the density of points again corresponds to a greater degree of stringy naturalness.
The defining feature of mirage mediation\cite{Choi:2005ge,Falkowski:2005ck} SUSY breaking models is they feature
comparable anomaly-mediated and gravity-mediated contributions to soft SUSY breaking terms, where the relative contributions are set by a parameter $\alpha$.
In particular, the unified gaugino masses typically expected at a high scale 
$Q=m_{\rm GUT}$ in gravity-mediation
are off-set by anomaly-mediated SUSY breaking (AMSB) contributions which are proportional to the gauge group beta functions. 
The gaugino mass running from $m_{\rm GUT}$ on down then compensates for
the AMSB off-set so that the gaugino masses unify at an intermediate scale 
\be
\mu_{\rm mir}=m_{\rm GUT}\cdot e^{(-8\pi^2/\alpha)}
\ee
where $m_{\rm GUT}\simeq 2\times 10^{16}$ GeV is where the MSSM gauge couplings unify.
Thus, the gaugino masses {\it appear} to unify at an intermediate (mirage) scale
which has no physical significance other than to stipulate the ratio of AMSB 
vs. universal gaugino mass boundary conditions.

The parameter space of the nGMM$^\prime$ model is given by\cite{Baer:2016hfa}
\be
\alpha,\ m_{3/2},\ c_m,\ c_{m3},\ a_3,\ \tan\beta,\ \mu,\ m_A\ \ \ (\text{nGMM}^\prime)
\ee
where $m_{3/2}$ is the gravitino mass (setting the overall scale of AMSB soft terms), 
$c_m$ ($c_{m3}$) set the scale of  first/second (third) generation AMSB bulk contributions, $a_3$ sets the scale of AMSB trilinear soft terms and
$\tan\beta$, $\mu$ and $m_A$ are the usual weak-scale MSSM parameters.
In original mirage-mediation (MM) models arising from orbifold compactifications, the 
$c_m$, $c_{m3}$ and $a_3$ parameters assumed discrete values, but in more general
and realistic Calabi-Yau flux compactifications these may be continuous.
(There are also $c_{H_u}$ and $c_{H_d}$ bulk parameters but these have been traded for
the more convenient weak scale MSSM parameters\cite{Baer:2016hfa}.)

The plot in Fig.~\ref{fig:GMM} shows the higgsino discovery plane with 
dots with density according to stringy naturalness as given by a 
landscape scan over the nGMM$^\prime$ model\cite{Baer:2019tee}.
In the plot, we adopt $m_{3/2}=20$ TeV, $m_A=2$ TeV, $\tan\beta =10$ with
$c_m=c_{m3}=((5~\mathrm{TeV})/\alpha M_s)^2$ (which sets the scalar masses near 5 TeV as in the
NUHM2 Fig.~\ref{fig:NUHM2}) and $M_s=m_{3/2}/(16\pi^2)$ and $a_3=1.6\sqrt{c_m}$.
We scan the landscape with a linear draw on $m_{1/2}^{MM}\equiv \alpha M_s$  
and let $\mu$ vary linearly from $100-360$ GeV. From the plot, we see that the
most stringy natural region has a mass splitting $\Delta m^0\equiv m_{\tchi_2^0}-m_{\tchi_1^0}\sim 4-5$ GeV. Smaller mass splittings require larger gaugino masses
which then lead to too large a weak scale beyond the ABDS window.
We also show the same ATLAS and CMS Run 2 exclusions (brown and black solid curves) and projected HL-LHC reaches as in Fig. \ref{fig:NUHM2}.

We compare against the $e^+e^-$ Higgs factory reach via EWPO via the green and purple contours (for $\Delta m_W =2.5$ and 0.5 MeV) and $\Delta\sin^2\theta_{\rm eff}=\pm 1.7\times 10^{-6}$ (magenta contour). 
We see that given enough precision on $\Delta m_W$, 
where FCC-ee is expected to probe to $\Delta m_W\sim 0.24$ MeV,
then the entire natural parameter space can be probed.
But the bulk region between the two magenta curves should give only
tiny deviations in $\Delta\sin^2\theta_{\rm eff}$, even beyond the
sensitivity of the FCC-ee Higgs factory.

\begin{figure}[tbp]
\begin{center}
 \includegraphics[clip, width = 0.8 \textwidth]{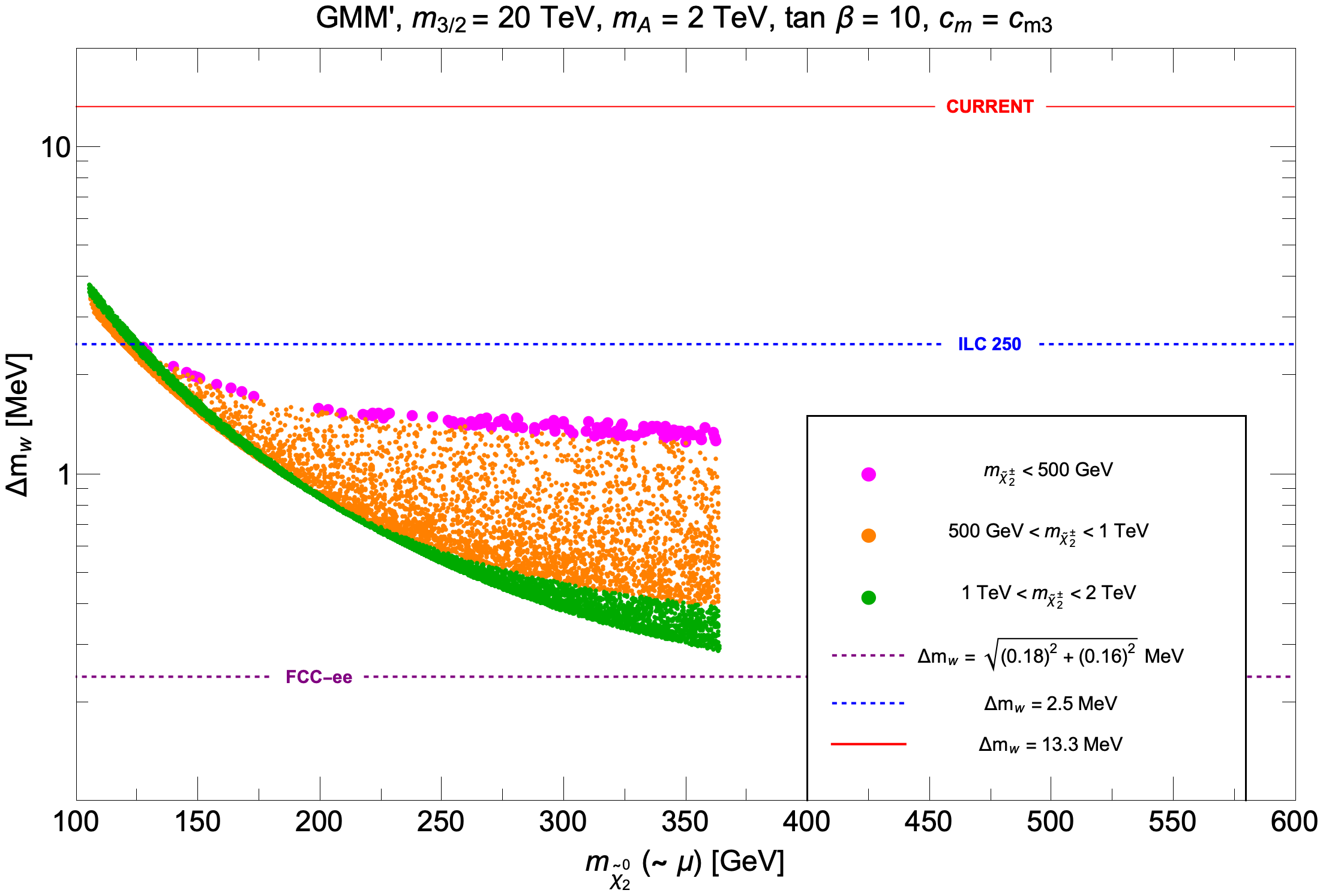}\\
\includegraphics[clip, width = 0.8 \textwidth]{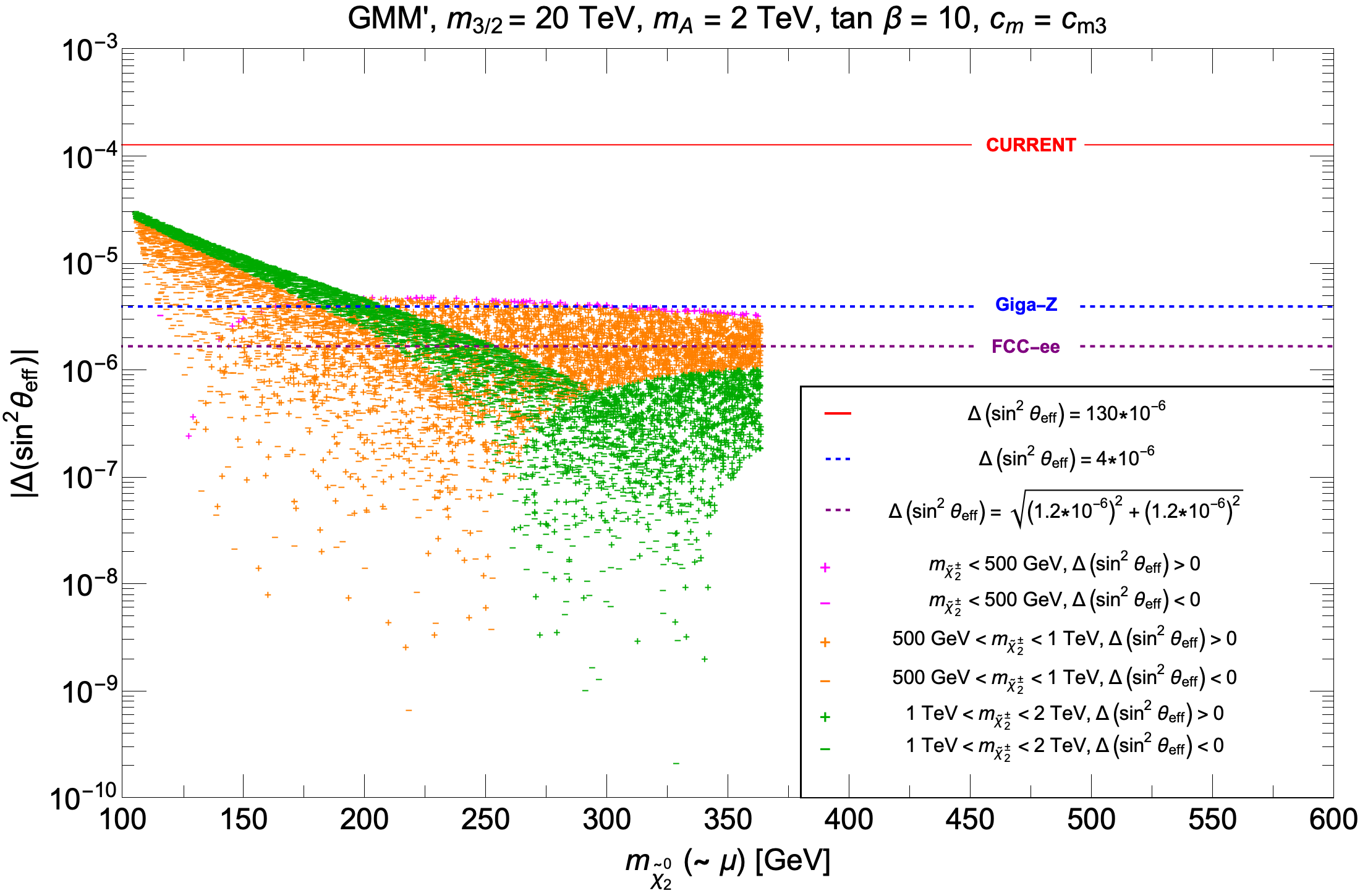}
 \caption{Plot of {\it a}) $\Delta m_W$ and {\it b}) $\Delta\sin^2\theta_{\rm eff}$ 
 vs. $m_{\tchi_2^0}$ from a scan over parameter space of the GMM$^\prime$ model with parameters as shown.
 }
\label{fig:gmm_scan}
\end{center}
\end{figure}

In Fig. \ref{fig:gmm_scan} we plot the same landscape theory dots but in the
{\it a}) $\Delta m_W$ and {\it b}) $\Delta\sin^2\theta_{\rm eff}$ vs. $m_{\tchi_2^0}$
(higgsino mass) planes. 
The color-coded dots are as in Fig. \ref{fig:nuhm_scan}.
We see again that at small $\mu$ and small $m(\text{wino})$, the expected deviations are
largest while for larger wino masses (preferred by the string landscape), 
the deviations decouple and are much smaller. 
While the entire set of points is accessible to FCC-ee via $\Delta m_W$, only a subset of points would give measurable deviations to FCC-ee via
$\Delta \sin^2\theta_{\rm eff}$.

\section{Summary and conclusions}
\label{sec:conclude}

In this paper, we have used the results of Ref.~\cite{Nagata:2025ycf}---which calculate the supersymmetric contributions to EWPOs $\Delta m_W$ and $\Delta\sin^2\theta_{\rm eff}$---to present corresponding results from two well-motivated
natural SUSY models (NUHM2 and nGMM) in the higgsino discovery plane,
$\Delta m^0$ vs. $m_{\tchi_2^0}$. The virtue of doing so allows us to 
first compare the reach of LHC, HL-LHC and various $e^+e^-$ collider
options with the theoretically preferred regions of parameter space and also
compare the reach of various present and future collider facilities, one against another. The most difficult portion of the higgsino discovery plane to probe is
the region with larger higgsino and wino masses: the lower-right portion of the
higgsino discovery plane. This region could elude HL-LHC searches via soft 
opposite-sign dileptons plus jets plus $\eslt$, and also elude searches by ILC500,
which would directly probe higgsino masses below $\sim 250$ GeV. 
Higher energy ILC searches with $\sqrt{s}\agt 2m(\text{higgsino})$ would cover the entire 
light higgsino  parameter space. However, $e^+e^-$ colliders operating as
$Z$ or Higgs factories apparently could also probe the entire natural
SUSY parameter space via the expected deviations in EWPOs. While any EWPO deviation
would be compelling, it would nonetheless not be entirely indicative of the
source of new physics, be it SUSY or some other manifestation. 
Detected deviations in EWPOs at proposed $e^+e^-$ Higgs factories could set the motivation for other future facilities so that direct production of new states
of matter could take place which would encode exactly what sort of new physics one was observing.

\section*{Acknowledgments}

HB gratefully acknowledges support from the Avenir Foundation.
VB gratefully acknowledges support from the William F. Vilas estate.
The work of NN was supported in part by the Grant-in-Aid for Scientific Research C (No.~25K07314).  

\bibliography{epem}
\bibliographystyle{elsarticle-num}
\end{document}